\documentclass[prx,twocolumn,english,superscriptaddress,floatfix,longbibliography]{revtex4-2} 

\usepackage{enumerate}
\usepackage{amsmath,amssymb,amsfonts}
\usepackage{hyperref,graphicx}
\usepackage{xcolor}
\graphicspath{{./Figures/}}
\hypersetup{
    colorlinks=true,
    citecolor=blue,
    linkcolor=magenta,
    urlcolor=magenta,
}

\usepackage{braket}
\usepackage{comment}
\usepackage{algorithmic}
\usepackage{algorithm}
\usepackage{overpic}
\usepackage{latexsym}
\usepackage[version=3]{mhchem}
\usepackage{tikz}
\usetikzlibrary{quantikz}

\usepackage{tabularx}    
\usepackage{array}       
\usepackage{booktabs}   
\usepackage{siunitx}    
\usepackage{adjustbox}
\usepackage[caption=false]{subfig}
\usepackage{multirow}
\usepackage{placeins}
\usepackage{float}
\usepackage{capt-of}

\newcommand{\cas}[2]{\ensuremath{(#1\mathrm{e},#2\mathrm{o})}}

\makeatletter
\renewcommand*{\@opargbegintheorem}[3]{\trivlist
      \item[\hskip \labelsep{\bfseries #1\ #2}] \textbf{(#3)}\ \itshape}
\makeatother

\begin{document}

\title{QSCI-CMP: Quantum-Selected Configuration Interaction with \\ Chemically Motivated Preselection}

\author{Masahiko Kamoshita}
\email{u312861c@ecs.osaka-u.ac.jp}
\affiliation{Graduate School of Engineering Science, The University of Osaka, 1-3 Machikaneyama, Toyonaka, Osaka 560-8531, Japan}

\author{Kosuke Mitarai}
\email{mitarai.kosuke.es@osaka-u.ac.jp}
\affiliation{Graduate School of Engineering Science, The University of Osaka, 1-3 Machikaneyama, Toyonaka, Osaka 560-8531, Japan}
\affiliation{Center for Quantum Information and Quantum Biology, The University of Osaka, 1-2 Machikaneyama, Toyonaka 560-0043, Japan}

\begin{abstract}
We present QSCI-CMP, a quantum-classical hybrid algorithm for molecular ground-state calculations that reduces both
the query count and the gate count of sample-based quantum diagonalization with amplitude amplification (SQD-AA).
SQD-AA mitigates the measurement bottleneck of quantum-selected configuration interaction (QSCI) by amplifying
the basis states that have not yet been measured.
Its oracle, however, specifies the measured states by listing them one by one, so its gate count grows with the number
of collected states. Moreover, the quantum resources are spent even on states whose importance is evident from chemical
knowledge, such as low-order excitations from the Hartree--Fock reference, which could be collected classically at the outset.
We therefore propose to fix such chemically trivial states in advance, include them in the diagonalization subspace
from the start, and exclude them from the amplification target, using a low-cost oracle that recognizes them
through the excitation level and the seniority number of each basis state.
We numerically demonstrate that QSCI-CMP reduces the query count and the gate count required to reach chemical accuracy
by up to approximately 68\% and 72\% relative to SQD-AA for 24-qubit systems.
The chemically trivial subspace is freely tunable within the classical computational budget.
A larger subspace shifts more work onto the classical solver and increases the reduction in quantum cost,
and when it captures the ground state sufficiently well, no quantum sampling is needed at all.
We also point out that a query-optimal iteration count known from the
analysis of quantum search further reduces the query count of both methods
by approximately 12\%.
\end{abstract}

\maketitle

\section{Introduction}
Quantum computing has recently attracted significant attention as a promising platform
for quantum chemistry calculations. In particular, the computation of ground-state
energies and the corresponding eigenvectors of molecular systems is expected to find
broad applications in fields such as drug discovery and materials design, and is
regarded as one of the principal targets where quantum computers may demonstrate an
advantage over classical methods.

Among the methods proposed for such ground-state calculations, quantum phase
estimation is the standard approach \cite{Kitaev1995, AspuruGuzik2005},
but it requires deep circuits with a large number of gates \cite{Reiher2017, Lee2021},
and its execution is expected to remain challenging in the near
future~\cite{Beverland2022}. As a more resource-efficient alternative,
quantum-selected configuration interaction (QSCI)~\cite{QSCI} has emerged.
In QSCI, an approximate ground state prepared on a quantum computer
is measured repeatedly in the computational basis, and the
Hamiltonian is diagonalized classically within the subspace spanned
by the sampled Slater determinants. We refer to this measurement
step as quantum sampling. Since its proposal, QSCI has been
extended in various directions~\cite{ADAPT-QSCI, TE-QSCI, HSB-QSCI, AFQMC-QSCI}, and
its practical viability has been demonstrated on current quantum devices, scaling up
to large-scale electronic-structure calculations~\cite{QSCI-IBM, HSB-QSCI,
Yoshioka2025Krylov, Yu2025SKQD, Shirakawa2025ClosedLoop}.
Its accuracy, however, is limited by the large number of measurements
required to capture determinants that carry only small amplitudes
in the prepared state yet are essential for recovering the
correlation energy~\cite{QSCI-Limitations}.

To mitigate this sampling bottleneck, sample-based quantum diagonalization with amplitude amplification (SQD-AA) has been proposed~\cite{SQD-AA}.
SQD-AA repeatedly applies amplitude amplification to suppress the amplitudes of the
determinants that have already been measured, so that the remaining determinants
become more likely to be observed, and thereby reduces the measurement cost of QSCI by
orders of magnitude. Yet SQD-AA leaves room for improvement in two aspects. First, the oracle
used in the amplification becomes a substantial part of the circuit as the iteration
proceeds. It specifies the already-measured determinants by listing them one by one,
so its gate count grows in proportion to the number of collected determinants.
Second, part of the quantum resources is spent on determinants that need not be
collected on a quantum computer in the first place.
In quantum chemistry the character of many important determinants
is known beforehand. Low-order excitations from the Hartree--Fock reference, for
instance, are almost certainly relevant, and such determinants can be enumerated
classically at the outset. SQD-AA nevertheless expends quantum resources on these
determinants, using measurements to collect them and oracle gates to exclude them
from subsequent rounds of amplification.

In this work, we propose QSCI-CMP, quantum-selected configuration interaction with chemically motivated preselection.
The idea is to treat the determinants whose importance is evident from chemical knowledge, which we call chemically
trivial, as if they had already been sampled.
They are fixed before any quantum sampling, entering the diagonalization subspace from the outset and
dropping out of the amplification target together with the actually measured determinants.
Quantum sampling and amplification are thereby reserved for the determinants whose importance cannot be foreseen classically.
The chemically trivial subspace is a free parameter of the method, bounded only by the cost of its classical diagonalization.
A larger subspace moves more of the computational burden to the classical side, and for systems whose important
determinants are entirely predictable, the method reduces to a purely classical diagonalization.

As a concrete choice of the chemically trivial subspace, we adopt determinant spaces
characterized by two structural quantities of each determinant, the excitation level
and the seniority number, i.e., the number of unpaired electrons. Both quantities are
read off directly from the occupation pattern of a determinant, and classical
electronic-structure theory has long used them to identify important determinants. Low
excitation levels dominate dynamical correlation~\cite{Helgaker2000}, while
low-seniority determinants capture the bulk of static correlation in
closed-shell-dominated systems~\cite{Bytautas2011, Bytautas2015}. In particular, the
hierarchy configuration interaction (hCI) framework~\cite{Kossoski2022} combines the
two quantities into a single hierarchy number and provides a family of determinant
spaces whose classical diagonalization cost grows systematically along the hierarchy,
which we use as preselection thresholds. QSCI-CMP is not tied to this choice, as any set
of determinants that can be enumerated classically may serve as the chemically
trivial subspace.

Our contributions beyond the framework itself are as follows. First, we construct an
oracle that identifies the chemically trivial subspace by computing the excitation
level and the seniority number in superposition, rather than by listing its elements,
so that its gate count is linear in the number of qubits and independent of the size
of the subspace. Second, we numerically demonstrate on hydrogen chains, \ce{H2O}, and
\ce{Cr2} that QSCI-CMP attains accuracy comparable to SQD-AA at a lower cost. For
24-qubit systems, the query count and the gate count required to reach chemical
accuracy of $1.6\,\mathrm{mHa}$ are reduced by up to approximately 68\% and 72\%, respectively, and when the
chemically trivial subspace captures the ground state sufficiently well, as for
\ce{Cr2}, it alone reaches chemical accuracy without any quantum sampling.

The remainder of this paper is organized as follows. Section~\ref{sec:background}
reviews QSCI, amplitude amplification, and SQD-AA. Section~\ref{sec:qscicmp} describes
the QSCI-CMP framework and the construction of its oracle.
Section~\ref{Experiments} presents the numerical experiments, and
Section~\ref{conclusion} concludes the paper.

\section{Background}\label{sec:background}

In this section, we review QSCI, amplitude amplification, and SQD-AA, and
discuss the limitations of SQD-AA that motivate this work.

\subsection{Quantum-Selected Configuration Interaction}\label{sec:qsci}

QSCI \cite{QSCI} is a hybrid quantum-classical algorithm for computing the
ground-state energy and the corresponding eigenvector of an electronic
Hamiltonian $\hat{H}$. The algorithm consists of three main steps: (i)
preparation of an approximate ground state $|\psi\rangle$ on a quantum
computer, (ii) sampling of the computational basis states by repeated
measurements of $|\psi\rangle$, and (iii) classical diagonalization of
$\hat{H}$ in the subspace spanned by the sampled basis states.

Let $\{|x\rangle\}_{x \in \{0,1\}^n}$ denote the computational basis states on
$n$ qubits, where each bit string $x$ labels a Slater determinant. Given an
approximate ground state
\begin{equation}
    |\psi\rangle = \sum_{x \in \{0,1\}^n} c_x |x\rangle,
    \label{eq:input_state}
\end{equation}
projective measurements of $|\psi\rangle$ in the computational basis sample
each $|x\rangle$ with probability $|c_x|^2$. We denote the set of sampled
basis states by $\mathcal{S}$. The ground-state energy is then
estimated by the lowest eigenvalue of the projected Hamiltonian
\begin{equation}
    \hat{H}_{\mathcal{S}} = P_{\mathcal{S}} \hat{H} P_{\mathcal{S}},
    \qquad
    P_{\mathcal{S}} = \sum_{x \in \mathcal{S}} |x\rangle\langle x|,
    \label{eq:projected_hamiltonian}
\end{equation}
computed on a classical computer.

The accuracy of QSCI depends on whether $\mathcal{S}$ contains the basis
states that contribute significantly to the true ground state. Sampling the
basis states with small amplitudes $|c_x|^2$ requires a large number of
measurements, and this constitutes the central bottleneck of QSCI
\cite{QSCI-Limitations}.

\subsection{Amplitude Amplification}\label{sec:aa}

Amplitude amplification \cite{Grover, BHMT} enhances the probability of
measuring a designated subset of basis states. Let
$|\psi\rangle = U|0\rangle^{\otimes n}$ be the state prepared by
a unitary $U$, and let $\mathcal{A} \subset \{0,1\}^n$ be a set
of basis states whose amplitudes are to be suppressed. Amplitude amplification
repeatedly applies the operator
\begin{equation}
    \hat Q = \hat O_\psi \hat O_{\mathcal{A}},
    \qquad
    \hat O_{\mathcal{A}} = I - 2 \sum_{x \in \mathcal{A}} |x\rangle\langle x|,
    \label{eq:aa_operator}
\end{equation}
where $\hat O_{\mathcal{A}}$ flips the sign of the amplitudes of the states in
$\mathcal{A}$, and $\hat O_\psi = I - 2|\psi\rangle\langle\psi|$ is a
reflection about $|\psi\rangle$, implemented as
$U R_0 U^{\dagger}$ with
$R_0 = I - 2\,(|0\rangle\langle 0|)^{\otimes n}$.

To see the effect of $\hat Q$, we decompose the initial state as
$|\psi\rangle = \sin\theta \, |\psi_{\mathrm{tgt}}\rangle
             + \cos\theta \, |\psi_{\perp}\rangle$,
where $|\psi_{\mathrm{tgt}}\rangle$ and $|\psi_{\perp}\rangle$ are the
normalized projections of $|\psi\rangle$ onto the subspaces spanned by $\overline{\mathcal{A}}$ and by $\mathcal{A}$, respectively,
and $\sin\theta = ( \sum_{x \notin \mathcal{A}} |c_x|^2 )^{1/2}$ with
$\theta \in (0,\pi/2)$. Each application of $\hat Q$ rotates the state by an
angle $2\theta$ in the plane spanned by these two states, so that after $k$
applications,
\begin{equation}
    \hat Q^k |\psi\rangle
      = \sin\bigl((2k+1)\theta\bigr) |\psi_{\mathrm{tgt}}\rangle
      + \cos\bigl((2k+1)\theta\bigr) |\psi_{\perp}\rangle,
    \label{eq:aa_iteration}
\end{equation}
and the probability of observing a target state,
$p(k) = \sin^2\bigl((2k+1)\theta\bigr)$, is maximized at
$k_{\mathrm{opt}} \approx \pi/(4\theta)$, where it becomes close to unity.

Preparing and measuring $\hat Q^k |\psi\rangle$ uses the state-preparation
unitary $2k+1$ times, once for the initial preparation and twice within each
$\hat Q$. 
The cost of $\hat O_{\mathcal{A}}$ depends on the structure of $\mathcal{A}$.
In the standard implementation, each element is specified by one
multi-controlled phase gate whose gate count is linear in the number of
qubits \cite{Barenco1995, Maslov2016}.

\subsection{Sample-Based Quantum Diagonalization with Amplitude
Amplification}\label{sec:sqdaa}
To mitigate the sampling bottleneck of QSCI, SQD-AA~\cite{SQD-AA}
suppresses the amplitudes of the basis states that have already been
measured, so that the remaining states become more likely to be
observed. Here the input state $\ket{\psi}$ is the approximate ground
state of Eq.~(\ref{eq:input_state}), and we write $U_{\mathrm{GS}}$
for the unitary that prepares it, $U = U_{\mathrm{GS}}$ in the
notation of Sec.~\ref{sec:aa}. Let $\mathcal{M} \subseteq \{0,1\}^n$
denote the set of measured basis states. SQD-AA chooses
$\mathcal{A} = \mathcal{M}$ in Eq.~(\ref{eq:aa_operator}),
\begin{equation}
    \hat O_{\mathcal{M}} = I - 2 \sum_{x \in \mathcal{M}} |x\rangle\langle x|,
    \label{eq:sqdaa_oracle}
\end{equation}
so that the target subspace is spanned by the unmeasured states
$\overline{\mathcal{M}} = \{0,1\}^n \setminus \mathcal{M}$.

The algorithm proceeds iteratively. Starting from $\mathcal{M} = \emptyset$,
the following steps are repeated:
\begin{enumerate}
    \item Construct the oracle $\hat O_{\mathcal{M}}$ from the current set
    $\mathcal{M}$.
    \item Apply $k$ rounds of amplitude amplification to $|\psi\rangle$.
    \item Perform $N_s$ projective measurements on the amplified state, and
    add the newly observed basis states to $\mathcal{M}$.
\end{enumerate}
The iteration count is chosen as $k \approx \pi/(4\theta)$ following
Sec.~\ref{sec:aa}, where $\theta$ is now determined by the weight of
the unmeasured states through
$\sin^2\theta = \sum_{x\notin\mathcal{M}}|c_x|^2$. This weight is not
known a priori, and in the original algorithm the $N_s$ measurements
of each iteration also serve to estimate its complement
$\sum_{x\in\mathcal{M}}|c_x|^2$, the weight of the measured states
\cite{SQD-AA}. In our resource analysis below, we assume $\theta$ to
be known when setting $k$. 
After a sufficient number of iterations, $\mathcal{M}$ contains basis states whose amplitudes in $|\psi\rangle$ are
small, and is used as the subspace $\mathcal{S}$ for the classical diagonalization of QSCI.

We remark on the choice of $k$. Since the algorithm samples repeatedly
from the amplified state, the relevant cost is the total number of
queries rather than the success probability $p(k)$ of each
measurement. The choice $k \approx \pi/(4\theta)$ maximizes only
$p(k)$, and the choice that minimizes the expected total query count
is known from the analysis of quantum search~\cite{BBHT}.
It stops the amplification slightly before
$p(k)$ reaches unity, at $(2k+1)\theta \approx 1.17$ where $p \approx 0.84$,
and reduces the expected total query count by approximately $12\%$,
independent of $\theta$ and of the target set. This applies to SQD-AA and
QSCI-CMP alike. In this work we follow the conventional choice for direct
comparability with Ref.~\cite{SQD-AA}.

The oracle $\hat O_{\mathcal{M}}$ leaves room for improvement in two respects.
First, it specifies the measured states by listing them one by one, one multi-controlled phase gate per element, so its gate count grows in
proportion to $|\mathcal{M}|$ and becomes a substantial part of the circuit as the iteration proceeds. We quantify this in Sec.~\ref{sec:meas_reduction}.
Second, as discussed in the introduction, part of $\mathcal{M}$, such as the low-order excitations from the Hartree--Fock reference, is known
to be important beforehand and can be enumerated classically, yet SQD-AA spends measurements to collect it and oracle gates to suppress it afterwards.
In the next section, we propose a method that addresses both points by incorporating this chemical knowledge into the oracle.

\section{QSCI-CMP}\label{sec:qscicmp}

\subsection{Framework}\label{sec:overview}

The central idea of QSCI-CMP is to replace the SQD-AA oracle
$\hat O_{\mathcal{M}}$ with an oracle that additionally suppresses a
chemically trivial subspace $\mathcal{T} \subset \{0,1\}^n$ fixed before any
quantum sampling,
\begin{equation}
    \hat O_{\mathrm{CMP}}
      = I - 2\sum_{x\in\mathcal{M}\cup\mathcal{T}}|x\rangle\langle x|,
    \label{eq:cmp_oracle}
\end{equation}
that is, the choice $\mathcal{A} = \mathcal{M}\cup\mathcal{T}$ in
Eq.~(\ref{eq:aa_operator}). The states in $\mathcal{T}$ are enumerated
classically and treated as if they had already been sampled. They are
included in the diagonalization subspace from the start and excluded from the
amplification target, so that the amplification acts only on the unmeasured,
chemically nontrivial basis states.

The workflow is that of SQD-AA with the oracle replaced. Starting from
$\mathcal{M} = \emptyset$, the following steps are repeated:
\begin{enumerate}
    \item Construct $\hat O_{\mathrm{CMP}}$ from the current set $\mathcal{M}$
    and the fixed subspace $\mathcal{T}$.
    \item Apply $k$ rounds of amplitude amplification to $|\psi\rangle$.
    \item Perform $N_s$ projective measurements on the amplified state, and
    add the newly observed basis states to $\mathcal{M}$.
\end{enumerate}
The iteration count is set as in Sec.~\ref{sec:sqdaa}, with
$\mathcal{M}$ replaced by $\mathcal{M}\cup\mathcal{T}$, and the same
set $\mathcal{M}\cup\mathcal{T}$ serves as the subspace for the
classical diagonalization.
In particular, QSCI-CMP starts from the accuracy obtained by
diagonalizing $\mathcal{T}$ alone, before any quantum sampling.

The subspace $\mathcal{T}$ is a free parameter of the method, and can
be taken as large as the available classical resources permit.
Enlarging it shifts more of the work onto the classical solver and, as we show in
Sec.~\ref{sec:meas_reduction}, increases the reduction in quantum
cost, and if $\mathcal{T}$ alone reaches the target accuracy, no
quantum sampling is performed at all. In the remainder of this section
we specify $\mathcal{T}$ through the excitation level and the
seniority number, and construct the corresponding oracle.

\subsection{Excitation Level and Seniority Number}\label{sec:exc_sen}

We characterize the chemically trivial subspace through two structural quantities of a Slater determinant, defined relative to the Hartree--Fock (HF) reference $|x_{\mathrm{HF}}\rangle$. 
Throughout, we use the Jordan--Wigner representation~\cite{JordanWigner1928}, in which the bit string $x \in \{0,1\}^n$ records the occupation of each spin orbital. 
This is not a restriction on the fermion-to-qubit mapping. For another mapping such as
Bravyi--Kitaev, the oracle constructed below is conjugated by the basis-conversion circuit to the Jordan--Wigner representation, a Clifford
circuit of $O(\log n)$ depth~\cite{Constantinides2025}, so the T-gate counts reported in this work are unchanged.

The excitation level $e(x)$ counts the number of single-particle excitations
required to transform $|x_{\mathrm{HF}}\rangle$ into $|x\rangle$,
\begin{equation}
    e(x) = \frac{1}{2} \sum_{i=0}^{n-1}
      \bigl( x_i \oplus x_{\mathrm{HF},i} \bigr),
    \label{eq:excitation}
\end{equation}
where $x_i$ and $x_{\mathrm{HF},i}$ denote the $i$-th bits of $x$ and
$x_{\mathrm{HF}}$, and $\oplus$ is the XOR operation, so that $e(x)$ is half
the Hamming distance between $x$ and $x_{\mathrm{HF}}$. The factor $1/2$
reflects that each excitation flips two bits, one occupied and one unoccupied.

The seniority number $\Omega(x)$ counts the singly occupied spatial orbitals.
Throughout, we index the bits of $x$ from zero and adopt the interleaved
ordering of the spin orbitals, in which the $\alpha$ and $\beta$ spin
orbitals of spatial orbital $p$ occupy the adjacent bits $x_{2p}$ and
$x_{2p+1}$, with $N_o = n/2$ the number of spatial orbitals. 
With this ordering, the seniority number reads
\begin{equation}
    \Omega(x) = \sum_{p=0}^{N_o-1} \bigl( x_{2p} \oplus x_{2p+1} \bigr).
    \label{eq:seniority}
\end{equation}

The two quantities are combined into the hierarchy number of the hCI
framework~\cite{Kossoski2022},
\begin{equation}
    h(x) = \frac{1}{2}\left( e(x) + \frac{\Omega(x)}{2} \right).
    \label{eq:hierarchy}
\end{equation}
Determinants with small $h$ are the chemically important ones, and the
threshold $h(x) \le h_{\max}$ defines a family of determinant spaces of
polynomially bounded size. 
The number of determinants with excitation level $e$ and seniority $\Omega$
scales as $\mathcal{O}(N_o^{\,e+\Omega/2})$. Since $e+\Omega/2 = 2h$, every
combination $(e,\Omega)$ on a given hierarchy $h$ contributes
$\mathcal{O}(N_o^{2h})$ determinants, so the space defined by $h \le h_{\max}$ has
$\mathcal{O}(N_o^{2h_{\max}})$ determinants~\cite{Kossoski2022}.
For example,
$h \le 2$ contains the full CISD space, since a double excitation has
$\Omega \le 4$ and hence $2e + \Omega \le 8$, and additionally admits
low-seniority higher excitations such as $(e,\Omega) = (3,2)$ and $(4,0)$,
which enter at the same $\mathcal{O}(N_o^{4})$ count; $h \le 2.5$ is the next
threshold in the hierarchy, with $\mathcal{O}(N_o^{5})$ determinants.

We define the chemically trivial subspace through a preselection
$\mathcal{R}$ in the $(e,\Omega)$ plane,
\begin{equation}
    \mathcal{T}(\mathcal{R})
      = \bigl\{ x \in \{0,1\}^n \mid (e(x), \Omega(x)) \in \mathcal{R} \bigr\}.
    \label{eq:trivial_subspace}
\end{equation}
The preselection $\mathcal{R}$ can be any subset of the $(e,\Omega)$ plane, and the
oracle of Sec.~\ref{sec:circuit} accommodates an arbitrary $\mathcal{R}$ at a
cost independent of $|\mathcal{T}|$. In this work we use thresholds on the
two quantities discussed above, $\mathcal{R} = \{e \le e_{\max}\}$ and
$\mathcal{R} = \{h \le h_{\max}\}$.

\subsection{Oracle Factorization}\label{sec:oracle_design}

The marked subspace of Eq.~(\ref{eq:cmp_oracle}) admits the decomposition
\begin{equation}
    \mathcal{M}\cup\mathcal{T}
      = \mathcal{T}\cup\bigl(\mathcal{M}\setminus\mathcal{T}\bigr),
    \label{eq:decomposition}
\end{equation}
where $\mathcal{T}$ is characterized by the structural conditions defining
$\mathcal{R}$, while $\mathcal{M}\setminus\mathcal{T}$ consists of explicitly
enumerated basis states. Since the two sets are disjoint, the corresponding
sign-flip operators commute, and $\hat O_{\mathrm{CMP}}$ factorizes as
\begin{equation}
    \hat O_{\mathrm{CMP}}
      = \hat O_{\mathcal{T}}\,\hat O_{\mathcal{M}\setminus\mathcal{T}},
    \label{eq:oracle_factorization}
\end{equation}
where
\begin{align}
    \hat O_{\mathcal{T}}
      &= I - 2\sum_{x\in\mathcal{T}}|x\rangle\langle x|, \label{eq:O_T}\\
    \hat O_{\mathcal{M}\setminus\mathcal{T}}
      &= I - 2\sum_{x\in\mathcal{M}\setminus\mathcal{T}}
        |x\rangle\langle x|. \label{eq:O_M_minus_T}
\end{align}
The crucial difference from the SQD-AA oracle lies in $\hat O_{\mathcal{T}}$.
As shown in Sec.~\ref{sec:circuit}, it is implemented at a cost independent of
$|\mathcal{T}|$ by evaluating the structural conditions in superposition,
whereas $\hat O_{\mathcal{M}\setminus\mathcal{T}}$ still lists its elements one by one. 
Enlarging $\mathcal{R}$ therefore moves configurations out of the enumerated remainder
$\mathcal{M}\setminus\mathcal{T}$ and into the structurally handled
$\mathcal{T}$, reducing the oracle cost, while removing more configurations from the amplification target. 
We examine this trade-off in Sec.~\ref{sec:meas_reduction}.

\subsection{Circuit Construction}\label{sec:circuit}

\begin{figure*}[t]
  \centering
  \subfloat[Excitation threshold]{%
    \resizebox{0.48\textwidth}{!}{%
    \begin{quantikz}[column sep=0.45cm, row sep={1.0cm,between origins}]
      \lstick{$\ket{x}_{\mathrm{sys}}$} & \qwbundle{n}
        & \gate{U_{\mathrm{HF}}} & \gate[3]{U_{\mathrm{cnt}}} & \qw & \gate[3]{\tilde{U}_{\mathrm{cnt}}^{\dagger}}
        & \gate{U_{\mathrm{HF}}^{\dagger}} & \qw \\
      \lstick{$\ket{0}_{\mathrm{tmp}}$} & \qwbundle{}
        & \qw & & \qw & & \qw & \qw \\
      \lstick{$\ket{0}_{W}$} & \qwbundle{\lceil\log_2(n{+}1)\rceil}
        & \qw & & \gate{Z_{\mathcal{R}}} & & \qw & \qw
    \end{quantikz}%
    }%
  }\hfill
  \subfloat[Hierarchy threshold]{%
    \resizebox{0.5\textwidth}{!}{%
    \begin{quantikz}[column sep=0.45cm, row sep={1.0cm,between origins}]
      \lstick{$\ket{x}_{\mathrm{sys}}$} & \qwbundle{n}
        & \gate{U_{\mathrm{HF}}} & \gate[2]{U_{\vee}} & \qw & \qw & \qw
        & \gate[2]{\tilde{U}_{\vee}^{\dagger}} & \gate{U_{\mathrm{HF}}^{\dagger}} & \qw \\
      \lstick{$\ket{0}_{\vee}$} & \qwbundle{n/2}
        & \qw & & \gate[3]{U_{\mathrm{cnt}}} & \qw & \gate[3]{\tilde{U}_{\mathrm{cnt}}^{\dagger}}
        & & \qw & \qw \\
      \lstick{$\ket{0}_{\mathrm{tmp}}$} & \qwbundle{}
        & \qw & \qw & & \qw & & \qw & \qw & \qw \\
      \lstick{$\ket{0}_{W}$} & \qwbundle{\lceil\log_2(n/2{+}1)\rceil}
        & \qw & \qw & & \gate{Z_{\mathcal{R}}} & & \qw & \qw & \qw
    \end{quantikz}%
    }%
  }
  \caption{Block-level circuits for $\hat O_{\mathcal{T}}$ at the two
  threshold preselections. In both cases the reference shift $U_{\mathrm{HF}}$
  forms $z=x\oplus x_{\mathrm{HF}}$, a single population-count network
  $U_{\mathrm{cnt}}$ writes the relevant integer onto an ancilla register, the
  phase flip $Z_{\mathcal{R}}$ flips the sign of the states inside the threshold,
  and the computation is uncomputed by measurement-based erasure at no T-gate cost
  (indicated by the tilde). (a) For an excitation threshold, $U_{\mathrm{cnt}}$
  counts the $n$ bits of $z$ into the Hamming weight $W=2e$. (b) For a hierarchy
  threshold, a layer $U_{\vee}$ first forms the $n/2$ pairwise ORs
  $z_{2p}\vee z_{2p+1}$, which $U_{\mathrm{cnt}}$ counts into $2h$. Each variant
  uses a single population count and costs $4n+\mathcal{O}(\log n)$ T gates.}
  \label{fig:blocks}
\end{figure*}

We now construct the two factors of Eq.~(\ref{eq:oracle_factorization})
explicitly. We quantify the circuit cost by the number of T gates, which
serves as the standard cost measure of fault-tolerant algorithms and
enables a direct comparison with the resource analysis of
SQD-AA~\cite{SQD-AA}. In fault-tolerant implementations, non-Clifford
gates such as the T gate require the dedicated preparation of magic
states, through distillation~\cite{BravyiKitaev2005} or
cultivation~\cite{Gidney2024Cultivation}. For the circuits considered
here, the T-gate count is moreover roughly proportional to the overall
gate count of both the oracle and the state-preparation circuit.

\subsubsection{Circuit for $\hat O_{\mathcal{T}}$}\label{sec:circuit_T}

The operator $\hat O_{\mathcal{T}}$ applies a phase flip to every $\ket{x}$
with $(e(x),\Omega(x))\in\mathcal{R}$. In all cases it first applies the
reference shift $U_{\mathrm{HF}}:\ket{x}\mapsto\ket{z}$,
$z\equiv x\oplus x_{\mathrm{HF}}$, a layer of Pauli-$X$ gates with no T cost,
computes onto an ancilla register an integer that determines membership in
$\mathcal{R}$, flips the phase accordingly, and uncomputes. The counting uses
the temporary logical-AND of Gidney~\cite{Gidney2018}, which costs four T
gates to compute and none to erase. Since the construction is out-of-place,
the uncomputation is T-free and the T-gate count is set entirely by the
compute pass. The cost depends on which integer must be computed, and hence on
$\mathcal{R}$. We summarize the three cases here with the circuits and
derivations given in Appendix~\ref{app:tcount}.

\paragraph{Excitation threshold.}
For $\mathcal{R}=\{e\le e_{\max}\}$, membership is fixed by the Hamming weight
$W=2e(x)$ of $z$, obtained by a single population count of its $n$ qubits,
shown in Fig.~\ref{fig:blocks}(a). The oracle costs
\begin{equation}
N_T^{e} \sim 4n
\label{eq:tcount_exc}
\end{equation}
T gates, with an additional $\mathcal{O}(\log n)$ T gates for the comparison $W\le 2e_{\max}$.

\paragraph{Hierarchy threshold.}
For $\mathcal{R}=\{h\le h_{\max}\}$, membership is fixed by the integer
$4h(x)=2e(x)+\Omega(x)$. For a closed-shell reference this equals
\begin{equation}
4h(x) = 2\sum_{p} \bigl(z_{2p}\vee z_{2p+1}\bigr),
\label{eq:fourh_or}
\end{equation}
the sum running over spatial-orbital pairs, so $2h$ is obtained by a single
population count of the $n/2$ pairwise ORs, shown in
Fig.~\ref{fig:blocks}(b). The oracle costs
\begin{equation}
N_T^{h} \sim 4n
\label{eq:tcount_hier}
\end{equation}
T gates. An open-shell reference is accommodated at the same cost, as shown in
Appendix~\ref{app:tcount}.

\paragraph{General preselection.}
For a general $\mathcal{R}$ that constrains $e$ and $\Omega$ independently,
both are computed as separate registers, requiring a half-adder layer followed
by two population counts. The oracle costs
\begin{equation}
N_T^{e,\Omega} \sim 6n,
\label{eq:tcount_general}
\end{equation}
where the additional $2n$ is the price of exposing $e$ and $\Omega$
separately rather than a single integer.

\subsubsection{Total Cost}\label{sec:t_count}

The operator $\hat O_{\mathcal{M}\setminus\mathcal{T}}$ specifies the
enumerated set $\mathcal{M}\setminus\mathcal{T}$ as a product of one
multi-controlled phase gate per basis state, exactly as in
SQD-AA~\cite{SQD-AA}. Each multi-controlled phase gate costs $\mathcal{O}(n)$
T gates~\cite{Barenco1995, Maslov2016}, so $\hat O_{\mathcal{M}\setminus\mathcal{T}}$
costs $\mathcal{O}(|\mathcal{M}\setminus\mathcal{T}|\cdot n)$ in total.

This construction leaves room for standard optimizations, such as the
cancellation of T gates between adjacent multi-controlled
gates~\cite{GidneyJones2021}, possibly aided by additional ancilla qubits.
These optimizations are available to SQD-AA and QSCI-CMP alike. Throughout
this work we do not apply them and count the per-element cost of both
methods.

Combining the two factors of Eq.~(\ref{eq:oracle_factorization}), and noting
that $\hat O_{\mathcal{T}}$ costs $\mathcal{O}(n)$ T gates in every case,
independent of $|\mathcal{T}|$, the total cost of the QSCI-CMP oracle is
\begin{equation}
N_T^{\mathrm{CMP}}
  = \mathcal{O}\bigl(|\mathcal{M}\setminus\mathcal{T}|\cdot n\bigr)
  + \mathcal{O}(n),
\label{eq:t_count_cmp}
\end{equation}
versus $\mathcal{O}(|\mathcal{M}|\cdot n)$ for SQD-AA. Once
$|\mathcal{M}\setminus\mathcal{T}|$ is appreciable the
$|\mathcal{T}|$-independent term is negligible, and the reduction in the
oracle cost relative to SQD-AA is governed by the fraction
$|\mathcal{M}\cap\mathcal{T}|/|\mathcal{M}|$ of measured configurations that
fall in $\mathcal{T}$, which grows as the threshold is enlarged.

\subsection{Comparison of the Reduction Rates}
\label{sec:cost_comparison}

QSCI-CMP reduces both the total gate count and the number of queries
to $U_{\mathrm{GS}}$ required to reach a given accuracy relative to
SQD-AA. To see how the two reductions relate to each other, we
consider the total gate count divided by the total number of queries
to $U_{\mathrm{GS}}$, that is, the average gate count per query to
$U_{\mathrm{GS}}$. Each application of
$\hat Q$ queries $U_{\mathrm{GS}}$ twice and the oracle once, so each
query to $U_{\mathrm{GS}}$ is accompanied by approximately $1/2$
queries to the oracle, the initial preparation of each amplified
state adding only a small correction. The average gate count per
query to $U_{\mathrm{GS}}$ is therefore
\begin{equation}
    C \approx G_{\mathrm{prep}} + \frac{1}{2}\,G_{\mathrm{oracle}},
    \label{eq:per_query_cost}
\end{equation}
where $G_{\mathrm{prep}}$ denotes the gate count of a single
$U_{\mathrm{GS}}$, and $G_{\mathrm{oracle}}$ denotes the gate count of a single oracle averaged over the queries to the oracle. 
The comparison of the two methods thus reduces to comparing their values of $C$.

The first term of $C$ is common to SQD-AA and QSCI-CMP. The two
methods therefore differ only through the second term, which is
proportional to the average of $|\mathcal{M}\setminus\mathcal{T}|$ for
QSCI-CMP and to that of $|\mathcal{M}|$ for SQD-AA. 
The total gate count to a certain accuracy is the product of
the number of queries to $U_{\mathrm{GS}}$ to that accuracy and $C$.
When QSCI-CMP reduces both factors, as observed in
Sec.~\ref{sec:meas_reduction}, the reduction rate of the total gate
count exceeds that of the number of queries to $U_{\mathrm{GS}}$.

Both reductions rest on the premise that the states in $\mathcal{T}$
carry appreciable weight in $\ket{\psi}$. This premise holds for
typical molecular systems, where the determinants of low excitation
level and low seniority dominate the ground state.

The difference between the two reduction rates depends on the ratio of
$G_{\mathrm{oracle}}$ to $G_{\mathrm{prep}}$. While $G_{\mathrm{prep}}$
is fixed by the ansatz, $G_{\mathrm{oracle}}$ grows as the iteration
proceeds. Whether it becomes comparable to $G_{\mathrm{prep}}$ before
chemical accuracy is reached cannot be foreseen, as it depends on the
amplitude distribution of the target system. When $G_{\mathrm{prep}}$
is small, as is the case for hardware-efficient state-preparation
ansatzes, $G_{\mathrm{oracle}}$ dominates $C$ and the difference grows
accordingly.
We assess this numerically in Sec.~\ref{sec:meas_reduction} through
the comparison of the query counts to $U_{\mathrm{GS}}$ and the
T-gate counts.

\section{Numerical Experiments}\label{Experiments}

\begin{figure*}[tbp]
  \centering
  \includegraphics[width=\textwidth]{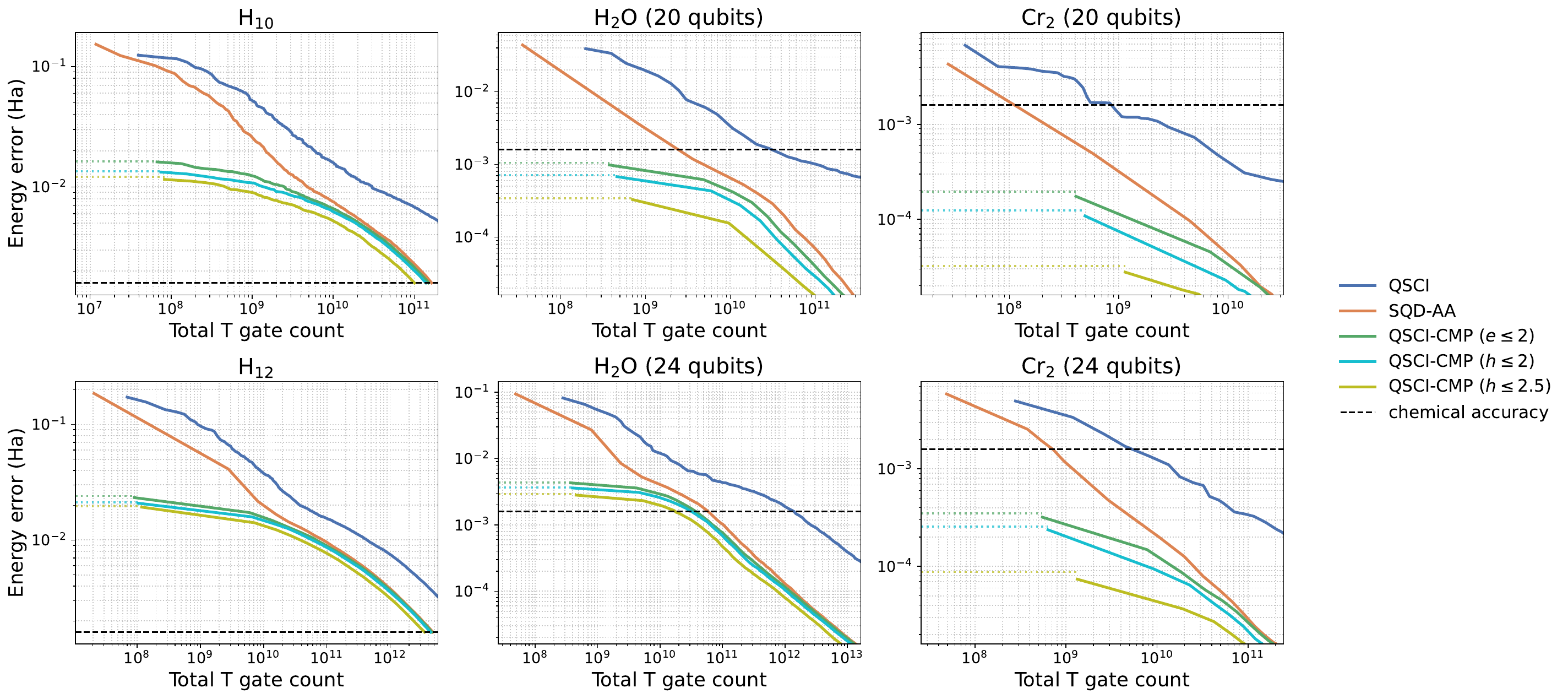}
  \caption{
    Energy error versus the cumulative T-gate count, comparing direct QSCI,
    SQD-AA, and QSCI-CMP at the three preselections $e\le2$, $h\le2$, and
    $h\le2.5$ for all six systems. The T-gate count combines the
    state-preparation cost at $L=N_o$ layers described in
    Sec.~\ref{sec:setup} and the oracle cost of each method. For direct
    QSCI, which uses no amplification, only the state-preparation cost
    enters. The top row shows the 20-qubit systems \ce{H10}, \ce{H2O}
    \cas{10}{10}, and \ce{Cr2} \cas{10}{10}, and the bottom row shows the
    24-qubit systems \ce{H12}, \ce{H2O} \cas{10}{12}, and \ce{Cr2}
    \cas{12}{12}. The dashed line marks chemical accuracy relative to the
    in-active-space reference. Colored dotted segments mark the initial
    accuracy of each QSCI-CMP variant at zero T gates, obtained by
    classically diagonalizing $\mathcal{T}(\mathcal{R})$ alone. Direct QSCI
    and SQD-AA have no such point, as they cannot identify important
    configurations prior to measurement.
  }
  \label{fig:tgate_all}
\end{figure*}

\begin{figure*}[tbp]
  \centering
  \includegraphics[width=\textwidth]{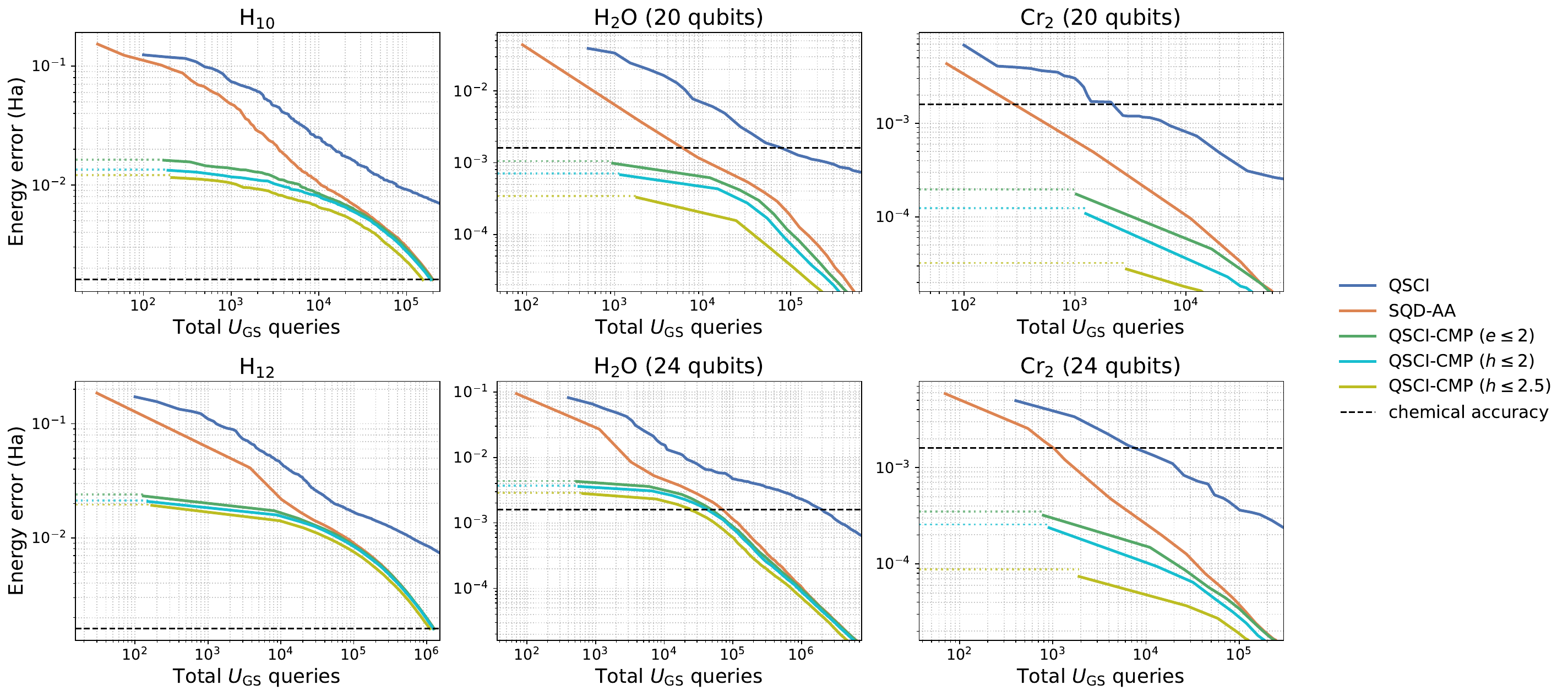}
  \caption{
    Same as Fig.~\ref{fig:tgate_all} but against the cumulative number of
    queries to the state-preparation unitary $U_{\mathrm{GS}}$.
  }
  \label{fig:query_all}
\end{figure*}

\subsection{Setup}\label{sec:setup}

We evaluate QSCI-CMP on three molecular systems chosen to span a range of
electron correlation and system size, namely linear hydrogen chains
\ce{H}$_N$, \ce{H2O}, and \ce{Cr2}. The hydrogen chains provide a tunable
model of strong correlation whose difficulty grows systematically with the
number of atoms~\cite{Motta2020}, \ce{H2O} is a closed-shell molecule
dominated by dynamical correlation, and the chromium dimer is a strongly
correlated transition-metal system that remains demanding even for
state-of-the-art classical methods~\cite{Larsson2022} and has served as a
benchmark for quantum-computational approaches to strong
correlation~\cite{Fomichev2024}. For the hydrogen chains we use the minimal
STO-3G basis and treat the full orbital space, following the original QSCI
benchmark setup~\cite{QSCI}, and for \ce{H2O} and \ce{Cr2} we use the
cc-pVDZ basis, following the basis-set choice standard in the QSCI
literature~\cite{SQD-AA, QSCI-IBM}. In all cases the orbitals are the
canonical RHF orbitals, and the active space is the frontier window of $N_o$
orbitals around the HOMO--LUMO gap. For each system the active space is
specified as $(N_e\mathrm{e}, N_o\mathrm{o})$. We treat the hydrogen chains
\ce{H10} and \ce{H12} in the full STO-3G space, i.e.\ \cas{10}{10} and
\cas{12}{12}, \ce{H2O} with \cas{10}{10} and \cas{10}{12}, and \ce{Cr2}
with \cas{10}{10} and \cas{12}{12}. The number of spin orbitals is
$n=2N_o$ in every case, ranging up to $n=24$.

The reference energy is the lowest eigenvalue obtained by exact
diagonalization within the chosen active space, and chemical accuracy is
defined relative to this in-active-space reference.
Note that accuracies reported here quantify convergence within the active
space and are not claims about full-orbital FCI or experimental energetics.

For the sampling experiments, the input state $\ket{\psi_{\mathrm{GS}}}$ is
the exact ground-state eigenvector obtained by diagonalization in the active
space. Drawing from the exact eigenvector isolates the effect of the oracle
design on sampling efficiency from the quality of any particular
state-preparation ansatz. The iteration count $k$ is set from the exact
weight of the marked states. Following
SQD-AA~\cite{SQD-AA}, we use $N_s=10$ for both methods, the value that gave
the largest cost reduction over most of its parameter study. All results
reported below are medians over 10 independent trials.

The chemically trivial subspace $\mathcal{T}(\mathcal{R})$ is controlled by
the preselection $\mathcal{R}$. We compare three choices, each a threshold
whose oracle costs $4n$ T gates as shown in Sec.~\ref{sec:circuit}. The first is an
excitation threshold $e(x)\le 2$, which masks all determinants up to double
excitations irrespective of seniority, and the other two are the hierarchy
thresholds $h\le 2$ and $h\le 2.5$ from the hCI framework~\cite{Kossoski2022}.
As discussed in Sec.~\ref{sec:exc_sen}, $h\le 2$ contains the CISD space
while additionally admitting low-seniority higher excitations, and $h\le 2.5$
is the next threshold in the hierarchy.

We compare the methods by two cost metrics, both of which approximate the
quantity of practical interest, the running time of the algorithm on a
quantum computer. The first is the T-gate count of the full circuit
including the oracle, which does so under a concrete cost model and is the
metric used in the resource analysis of SQD-AA~\cite{SQD-AA}. The T-gate
count, however, rests on the choice of the state-preparation ansatz, for
which no standard exists. The second metric is the number of queries to the
state-preparation unitary $U_{\mathrm{GS}}$, which is free of this choice. When the state-preparation circuit is far more expensive
than the oracle, the running time is approximately proportional to the
number of queries to $U_{\mathrm{GS}}$.

For the T-gate cost analysis, we assume that the unitary cluster Jastrow
(UCJ) ansatz~\cite{Matsuzawa2020, Motta2023} is expressive enough to prepare
$\ket{\psi_{\mathrm{GS}}}$ exactly, and count the T gates of the
state-preparation circuit following the cost model of SQD-AA~\cite{SQD-AA}.
In this model, the $L(\tfrac{5}{2}n^2-\tfrac{1}{2}n)$ single-qubit rotations
of an $L$-layer UCJ circuit are each synthesized over Clifford$+$T by a
repeat-until-success circuit~\cite{Bocharov2014}, with the synthesis accuracy
set by a total error budget of $\epsilon_{\mathrm{tot}}=10^{-4}$~\cite{Kivlichan2020}.
In the UCJ ansatz, each layer corresponds to one component of a low-rank
factorization of the two-body cluster amplitudes, whose rank scales linearly
with the number of spatial orbitals~\cite{Motta2021}, so a faithful circuit
uses $L=\mathcal{O}(N_o)$ layers, and we take $L=N_o$. This estimate applies
to the full UCJ form and upper-bounds the local unitary cluster Jastrow
(LUCJ) ansatz~\cite{Motta2023}, its hardware-efficient variant, which has
served as the state-preparation circuit in recent
demonstrations~\cite{QSCI-IBM}. The resulting T-gate count depends
only on the ansatz structure, i.e.\ the number of layers and the interaction
pattern on $n$ qubits, and not on the variational parameters. The oracle
cost is counted per element, with
one multi-controlled phase gate per enumerated configuration for both
methods and an additional $4n$ T gates per call for
$\hat O_{\mathcal{T}}$ in QSCI-CMP.

\subsection{T-gate and Query Reduction}\label{sec:meas_reduction}

Figure~\ref{fig:tgate_all} compares the energy error against the cumulative
T-gate count, which includes the oracle cost, for direct QSCI, SQD-AA, and
the QSCI-CMP variants on all six systems. The six systems fall into two
groups according to the initial error obtained by diagonalizing
$\mathcal{T}(\mathcal{R})$ alone, shown as the dotted segments at zero T
gates. For \ce{H2O} in the \cas{10}{10} active space and for \ce{Cr2} in
both active spaces, the initial error is already below chemical accuracy
for all three preselections, so QSCI-CMP performs no quantum sampling and
reduces to a purely classical diagonalization, whereas SQD-AA requires on
the order of $10^3$ to $10^4$ queries to $U_{\mathrm{GS}}$ for these
systems. Notably, both molecules served as benchmark systems of the
original SQD-AA proposal, in the same basis and active spaces~\cite{SQD-AA}.
For the remaining three systems, \ce{H10}, \ce{H12}, and \ce{H2O} in the
\cas{10}{12} active space, the initial error lies above chemical accuracy
and quantum sampling is required. Quantum resources are thus spent only
where the classical subspace falls short, and we focus on these three
systems in the following.

For all three systems, the QSCI-CMP variants reach chemical accuracy with
fewer T gates than SQD-AA, and the reduction grows as the preselection is
enlarged from $e\le2$ to $h\le2.5$. At $h\le2.5$, the T-gate count to
reach chemical accuracy is reduced by $40.2\%$, $26.7\%$, and $72.5\%$ for
\ce{H10}, \ce{H12}, and \ce{H2O} in the \cas{10}{12} active space,
respectively. Across the three systems, the reduction is larger where the
initial error is lower. The T-gate reduction combines the reduction of the
oracle, which replaces the enumerated configurations that fall in
$\mathcal{T}$ with the $|\mathcal{T}|$-independent $\hat O_{\mathcal{T}}$,
with the reduction of the number of queries to $U_{\mathrm{GS}}$, which we
examine next.

Figure~\ref{fig:query_all} shows the same comparison against the cumulative
number of queries to $U_{\mathrm{GS}}$. At $h\le2.5$, the query count to
reach chemical accuracy is reduced by $25.1\%$, $12.4\%$, and $68.4\%$ for
\ce{H10}, \ce{H12}, and \ce{H2O} in the \cas{10}{12} active space,
respectively. In each case, the relative reduction in the query count is
smaller than that in the T-gate count, as anticipated in
Sec.~\ref{sec:cost_comparison}, since the query count leaves out the
reduction of the oracle. The difference grows with the ratio of
$G_{\mathrm{oracle}}$ to $G_{\mathrm{prep}}$. At the iteration where
chemical accuracy is reached, the oracle accounts for approximately $55\%$
and $83\%$ of the gate count per query to $U_{\mathrm{GS}}$ in SQD-AA for
\ce{H10} and \ce{H12}, respectively. $G_{\mathrm{oracle}}$ thus grows to a
size comparable to $G_{\mathrm{prep}}$ before chemical accuracy is reached.
This quantifies the growth
of the oracle noted in Sec.~\ref{sec:sqdaa} and confirms that, for the
state-preparation circuits considered here, the oracle is a substantial
part of the total cost.

\section{Conclusion}\label{conclusion}

We proposed QSCI-CMP, a method that treats the basis states whose importance
is evident from chemical knowledge as if they had already been sampled.
They are fixed before any quantum sampling, placed in the diagonalization subspace from the beginning, and removed
from the amplification target of SQD-AA.
Quantum sampling and amplification are thereby reserved for the configurations
whose importance cannot be foreseen classically. As a concrete realization, we
characterized the chemically trivial subspace through the excitation level and
the seniority number, and constructed an oracle that evaluates these
structural conditions in superposition, at a T-gate count of $4n$ for an
excitation or hierarchy threshold and $6n$ for a general preselection, in
all cases independent of the size of the subspace.

We numerically demonstrated that QSCI-CMP attains accuracy comparable to
SQD-AA on hydrogen chains, \ce{H2O}, and \ce{Cr2}, at a lower cost. For
24-qubit systems, the query count and the T-gate count required to reach
chemical accuracy are reduced by up to approximately 68\% and 72\% relative
to SQD-AA.
The chemically trivial subspace is a tunable parameter, limited only by
the classical cost of diagonalizing it. Enlarging it offloads more of the computation to the classical
solver and increases the reduction in quantum cost, and for systems already within classical reach the method
reduces to a purely classical diagonalization.

Several limitations remain. The sampling experiments draw from the exact
ground-state eigenvector and thus do not account for the finite fidelity of a
realistic state-preparation ansatz, and the reported accuracies are defined
relative to the in-active-space reference rather than the full-orbital or
experimental energetics. The oracle $\hat O_{\mathcal{M}\setminus\mathcal{T}}$
is evaluated with the per-element multi-controlled-phase construction, though
a more efficient implementation would change its cost and the size of the
reported reductions.

These limitations point to several directions for future work. Extending the
sampling to states prepared by an explicit ansatz of finite fidelity would
clarify the behavior under realistic state preparation. Applying QSCI-CMP to
more strongly correlated regimes, such as stretched geometries and larger
active spaces, is another natural step. More broadly, the excitation level and
the seniority number are only one way to specify the chemically trivial
subspace; any classically enumerable set of determinants can serve, and
incorporating richer chemical knowledge into the preselection is a promising
avenue toward practical quantum chemistry algorithms.

\section*{Acknowledgement}
This work is supported by JST COI-NEXT Grant No. JPMJPF2014, JST Moonshot R\&D Grant No. JPMJMS256J, MEXT Quantum Leap Flagship Program (MEXT Q-LEAP) Grant No. JPMXS0120319794, and JST NEXUS Grant No. JPMJNX26C6.
K.M. is supported by JSPS KAKENHI Grant No. 23H03819, 24K16980, and JST CREST Grant Number JPMJCR24I4.
M.K. is supported by JST SPRING under Grant Number JPMJSP2138.

\appendix

\section{Circuit and T-gate Count of \texorpdfstring{$\hat O_{\mathcal{T}}$}{O\_T}}
\label{app:tcount}

This appendix details the construction of $\hat O_{\mathcal{T}}$ and derives
the T-gate counts quoted in Sec.~\ref{sec:circuit}. We show that the
excitation level $e$, the seniority $\Omega$, and the hierarchy number $h$ of
a computational basis state are all obtained as Hamming weights of simple
two-bit functions of the reference-shifted string. The oracle circuits and
their T-gate counts then follow from known fault-tolerant constructions,
chiefly the temporary logical-AND of Ref.~\cite{Gidney2018}. We first review
these constructions, and then construct the oracles for the excitation
threshold, the hierarchy threshold, and the general preselection in turn.

Throughout, $n$ denotes the number of spin orbitals, $N_o=n/2$ the number of
spatial orbitals, and $z=x\oplus x_{\mathrm{HF}}$ the reference-shifted
string produced by $U_{\mathrm{HF}}$. Under the interleaved ordering, the two
spin orbitals of the spatial orbital $p$ occupy the adjacent bits
$(z_{2p},z_{2p+1})$.

\subsection{Building Blocks: Hamming-Weight Computation and Threshold Comparison}
\label{app:blocks}

Every variant of $\hat O_{\mathcal{T}}$ computes the Hamming weight of a
string of single-bit values on an ancilla register and then compares it with
a classical threshold. Both steps use the temporary
logical-AND of Ref.~\cite{Gidney2018}, which costs four T gates
and is later erased by an $X$-basis measurement and a classically controlled
Clifford correction at no T cost.

Reference~\cite{Gidney2018} describes how to compute the Hamming weight of
$m$ input bits into a $\lceil\log_2(m{+}1)\rceil$-bit
register by repeatedly applying an out-of-place full adder built from one
temporary logical-AND. Each adder reads three bits of equal binary weight, or
two when only two bits of that weight remain, and writes their sum bit and
carry bit onto fresh ancillas. The inputs and all intermediate bits are
preserved, so the entire computation is later erased by measurement-based
uncomputation without any T gates, and the T-gate count of the computation
is bounded by $4m$. We denote the resulting
network by $U_{\mathrm{cnt}}$. Writing $b=(b_0,\dots,b_{m-1})$ for the input
bits, it acts as
\begin{equation}
U_{\mathrm{cnt}}\,
\ket{b}\ket{0}_{\mathrm{tmp}}\ket{0}_{W}
=\ket{b}\ket{g(b)}_{\mathrm{tmp}}
\Bigl|\textstyle\sum_{i=0}^{m-1} b_i\Bigr\rangle_{W},
\label{eq:ucnt_action}
\end{equation}
where the register $W$ holds the binary representation of the sum and the
register tmp holds the intermediate bits, denoted collectively by $g(b)$. A
direct count shows that the network uses exactly $m-\mathrm{wt}_2(m)$ full
and half adders in total~\cite{MullerPreparata1975}, where
$\mathrm{wt}_2(m)$ is the number of ones in the binary representation of
$m$, and therefore
\begin{equation}
4\bigl(m-\mathrm{wt}_2(m)\bigr)
\label{eq:popcount_tcount}
\end{equation}
T gates. This number of ANDs is optimal. Computing the Hamming weight of $m$
bits using ANDs together with CNOT and $X$ gates requires at least
$m-\mathrm{wt}_2(m)$ ANDs by Theorem~7 of Ref.~\cite{BoyarPeralta2008}. An explicit diagram of
such a counting network, for $m=8$, is given
in Supplementary Figure~2 of Ref.~\cite{KanSymons2025}. Any other AND
or Toffoli implementation changes
only the per-AND coefficient of four and leaves the construction intact.

The threshold comparison is obtained from the ripple-carry adder of
Ref.~\cite{Cuccaro2004} by computing only the output carry of a subtraction,
with each carry produced by one temporary logical-AND. The
comparator writes the truth value of the comparison onto a flag qubit, a $Z$
gate on the flag applies the phase flip, and the comparator is then
uncomputed for free in the same way as $U_{\mathrm{cnt}}$. Acting on the
$\lceil\log_2(m{+}1)\rceil$-bit result register, the comparator adds
$\mathcal{O}(\log m)$ T gates. We use this bound whenever a threshold
comparison appears below.

\subsection{Excitation Threshold}

For an excitation threshold $\mathcal{R}=\{e\le e_{\max}\}$, the only quantity
required is the excitation level, which is determined by the Hamming weight
of $z$ through $W(z)=2e(x)$. The oracle is therefore a single Hamming-weight
computation on the $n$ bits of $z$, performed by the network of
Eq.~(\ref{eq:ucnt_action}) with $m=n$, followed by a phase flip on the states
with $W\le 2e_{\max}$. By Eq.~(\ref{eq:popcount_tcount}) the count costs
\begin{equation}
N_T^{e} = 4\bigl(n-\mathrm{wt}_2(n)\bigr) \sim 4n
\label{eq:tcount_e}
\end{equation}
T gates and is uncomputed for free. The threshold comparison acts on the
$\lceil\log_2(n{+}1)\rceil$-bit weight register and adds $\mathcal{O}(\log n)$
T gates by the comparator of Sec.~\ref{app:blocks}. No half-adder layer is
required, because the excitation level is determined by the Hamming weight of
$z$ alone.

\subsection{Hierarchy Threshold}

For a hierarchy threshold $\mathcal{R}=\{h\le h_{\max}\}$, membership is
decided by the integer $4h(x)=2e(x)+\Omega(x)$. We show that for a
closed-shell reference this integer is twice the Hamming weight of $N_o$
pairwise OR bits, and that an open-shell reference is accommodated at the
same cost by replacing the OR with an AND on the singly occupied pairs.

\subsubsection{Closed-Shell Reference}

For a closed-shell reference, every spatial-orbital pair of
$x_{\mathrm{HF}}$ is $(0,0)$ or $(1,1)$, so $z_{2p}=1$ exactly when the
occupation of the spin orbital $2p$ in $x$ differs from that in
$x_{\mathrm{HF}}$, and likewise for $z_{2p+1}$. We decompose
$4h(x)=2e(x)+\Omega(x)$ into the contributions of the individual spatial
orbitals. The spatial orbital $p$ adds $2$ to $2e(x)$ when the occupations of
both of its spin orbitals differ from the reference, $z_{2p}z_{2p+1}=11$, and
it then adds nothing to $\Omega(x)$ because the pair is either empty or
doubly occupied in $x$. It adds $1$ to $2e(x)$ and $1$ to $\Omega(x)$ when
exactly one occupation differs, $z_{2p}z_{2p+1}=01$ or $10$, because the pair
is then singly occupied in $x$. It adds nothing to either quantity when the
pair matches the reference, $z_{2p}z_{2p+1}=00$. In all three cases the
amount added to $4h(x)$ equals $2\,(z_{2p}\vee z_{2p+1})$. Summing over the
spatial orbitals,
\begin{equation}
4h(x) = 2e(x)+\Omega(x)
      = 2\sum_{p=0}^{N_o-1}\bigl(z_{2p}\vee z_{2p+1}\bigr).
\label{eq:fourh_or_app}
\end{equation}
Equivalently, $2h(x)$ is equal to the number of spatial orbitals in which the
occupation of at least one of the two spin orbitals differs between $x$ and
$x_{\mathrm{HF}}$.

The oracle therefore consists of a layer $U_{\vee}$ that writes the OR bits
$v_p=z_{2p}\vee z_{2p+1}$ for $p=0,\dots,N_o-1$ onto fresh ancillas, followed
by the Hamming-weight computation on the $v_p$ and the threshold comparison.
Each OR
bit is obtained with a single temporary logical-AND through the identity
$z_{2p}\vee z_{2p+1}=\neg(\neg z_{2p}\wedge\neg z_{2p+1})$, realized by
initializing the ancilla to $\ket{1}$ with an $X$ gate and flipping it with
the AND applied with both controls open, so $U_{\vee}$ uses $N_o=n/2$ ANDs.
The Hamming-weight computation on the $N_o$ OR bits uses
$N_o-\mathrm{wt}_2(N_o)$ ANDs by Eq.~(\ref{eq:popcount_tcount}), and after
the count the result register holds the integer $\sum_p v_p=2h(x)$. The
total cost is
\begin{align}
N_T^{h} &= \underbrace{4\cdot\tfrac{n}{2}}_{U_{\vee}}
        + \underbrace{4\Bigl(\tfrac{n}{2}-\mathrm{wt}_2\bigl(\tfrac{n}{2}\bigr)\Bigr)}_{U_{\mathrm{cnt}}}
        \nonumber\\
        &\sim 4n
\label{eq:tcount_h_app}
\end{align}
T gates, uncomputed for free. The comparison of the result register with the
threshold, which enforces $2h(x)\le 2h_{\max}$, adds $\mathcal{O}(\log n)$ T
gates by the comparator of Sec.~\ref{app:blocks}.

\subsubsection{Open-Shell Reference}

When the reference carries unpaired electrons, the singly occupied pairs of
$x_{\mathrm{HF}}$ are $(0,1)$ or $(1,0)$, and the contribution of such a pair
to $4h(x)$ is no longer $2\,(z_{2p}\vee z_{2p+1})$. A direct enumeration of
the four occupations $(x_{2p},x_{2p+1})$ shows that a singly occupied
reference pair adds
\begin{equation}
1 + 2\,(z_{2p}\wedge z_{2p+1})
\label{eq:pair_open}
\end{equation}
to $4h(x)$. For the occupations with $z_{2p}z_{2p+1}=00$, $01$, or $10$ the
pair adds $1$ in total, either as the unit of seniority it already carries in
the reference or as one unit of $2e(x)$ when one occupation differs. For the
occupation with $z_{2p}z_{2p+1}=11$, in which the unpaired electron has moved
to the opposite spin orbital of the same spatial orbital, the pair adds $3$,
two units of $2e(x)$ and one of $\Omega(x)$. The doubly occupied and empty
pairs of the reference add $2\,(z_{2p}\vee z_{2p+1})$ as in the closed-shell
case. Hence
\begin{equation}
4h(x) = \Omega_{\mathrm{ref}}
  + 2\!\!\sum_{p\in P_{\mathrm{c}}}\!\!(z_{2p}\vee z_{2p+1})
  + 2\!\!\sum_{p\in P_{\mathrm{o}}}\!\!(z_{2p}\wedge z_{2p+1}),
\label{eq:fourh_open}
\end{equation}
where $P_{\mathrm{o}}$ is the set of singly occupied reference pairs,
$P_{\mathrm{c}}$ is its complement, and
$\Omega_{\mathrm{ref}}=|P_{\mathrm{o}}|$ is the seniority of the reference.
Since the reference, and hence the partition into $P_{\mathrm{o}}$ and
$P_{\mathrm{c}}$, is fixed when the circuit is designed, the layer
$U_{\vee}$ is modified only on the pairs in $P_{\mathrm{o}}$, where the OR
is replaced by an AND at the same cost of one temporary logical-AND, and the
constant $\Omega_{\mathrm{ref}}$ is absorbed into the comparison threshold
at no gate cost. The Hamming-weight computation is unchanged, so the T-gate
count is
again $N_T^h\sim 4n$. The construction therefore extends to open-shell
references, including odd-electron systems, without additional cost.

\subsection{General Preselection}

\begin{figure*}[t]
  \centering
  \begin{quantikz}[column sep=0.45cm, row sep={1.0cm,between origins}]
    \lstick{$\ket{x}_{\mathrm{sys}}$} & \qwbundle{n}
      & \gate{U_{\mathrm{HF}}} & \gate[2]{U_{\mathrm{HA}}} & \qw & \qw & \qw & \qw & \qw
      & \gate[2]{\tilde{U}_{\mathrm{HA}}^{\dagger}} & \gate{U_{\mathrm{HF}}^{\dagger}} & \qw \\
    \lstick{$\ket{0}_{s,c}$} & \qwbundle{n}
      & \qw & & \gate[4]{U_{\mathrm{cnt}}} & \qw & \qw & \qw & \gate[4]{\tilde{U}_{\mathrm{cnt}}^{\dagger}}
      & & \qw & \qw \\
    \lstick{$\ket{0}_{\mathrm{tmp}}$} & \qwbundle{}
      & \qw & \qw & & \qw & \qw & \qw & & \qw & \qw & \qw \\
    \lstick{$\ket{0}_{W_s}$} & \qwbundle{}
      & \qw & \qw & & \gate[3]{U_W} & \gate[3]{Z_{\mathcal{R}}} & \gate[3]{\tilde{U}_W^{\dagger}} & & \qw & \qw & \qw \\
    \lstick{$\ket{0}_{W_c}$} & \qwbundle{}
      & \qw & \qw & & & & & \qw & \qw & \qw & \qw \\
    \lstick{$\ket{0}_{W}$} & \qwbundle{\lceil\log_2(n{+}1)\rceil}
      & \qw & \qw & \qw & & & & \qw & \qw & \qw & \qw
  \end{quantikz}
  \caption{Block-level circuit for the general-preselection oracle
  $\hat O_{\mathcal{T}}$. The half-adder layer $U_{\mathrm{HA}}$ of
  Eq.~(\ref{eq:uha_action}) writes the sum bits $\{s_p\}$ and carry bits
  $\{c_p\}$ onto $\ket{0}_{s,c}$. The out-of-place counting networks
  $U_{\mathrm{cnt}}$ of Eq.~(\ref{eq:ucnt_action}), built from the full
  adders of Ref.~\cite{Gidney2018}, use the workspace $\ket{0}_{\mathrm{tmp}}$
  to sum the $n/2$ bits of each group and write $W_s=\sum_p s_p$ onto
  $\ket{0}_{W_s}$ and $W_c=\sum_p c_p$ onto $\ket{0}_{W_c}$. For a
  closed-shell reference, $W_s$ equals the seniority $\Omega$. The combiner
  $U_W$ of Eq.~(\ref{eq:uw_action}) then forms $W=W_s+2W_c$ on $\ket{0}_{W}$,
  leaving $W_s$ intact. The phase flip $Z_{\mathcal{R}}$ of
  Eq.~(\ref{eq:zr_action}) reads the excitation level $e=W/2$ from
  $\ket{0}_{W}$ and the seniority $\Omega=W_s$ from $\ket{0}_{W_s}$, flipping
  the sign of exactly the configurations whose $(e,\Omega)$ lies in
  $\mathcal{R}$. The tilde on $\tilde{U}_W^{\dagger},
  \tilde{U}_{\mathrm{cnt}}^{\dagger}, \tilde{U}_{\mathrm{HA}}^{\dagger}$
  indicates measurement-based erasure rather than a plain inverse unitary.
  Each temporary logical-AND is removed by an $X$-basis measurement and a
  classically controlled Clifford correction, so the uncomputation uses no T
  gates. Only the phase imparted by $Z_{\mathcal{R}}$ remains.}
  \label{fig:circuit_cmp}
\end{figure*}

For a general preselection that constrains $e$ and $\Omega$ independently,
both quantities must be held in separate registers. The block-level circuit
is shown in Fig.~\ref{fig:circuit_cmp}. We state the construction for a
closed-shell reference and comment on the open-shell case at the end. This most general variant subsumes the two
thresholds above at a higher cost.

\paragraph{Half-adder layer $U_{\mathrm{HA}}$.}
The layer splits each pair $(z_{2p},z_{2p+1})$ into a sum bit and a carry
bit, acting as
\begin{align}
U_{\mathrm{HA}}\,\ket{z}\ket{0}^{\otimes n}
 &= \ket{z}\,\ket{s_0\cdots s_{N_o-1}}\,\ket{c_0\cdots c_{N_o-1}},
 \nonumber\\
s_p &= z_{2p}\oplus z_{2p+1},
\qquad
c_p = z_{2p}\wedge z_{2p+1}.
\label{eq:uha_action}
\end{align}
Each sum bit is written by two CNOTs at no T cost and each carry bit by one
temporary logical-AND, so the layer uses $N_o=n/2$ ANDs. The inputs $z$ are
preserved.

\paragraph{Hamming-weight computations.}
Two applications of the network of Eq.~(\ref{eq:ucnt_action}), each counting
$N_o$ bits with $N_o-\mathrm{wt}_2(N_o)$ ANDs, write the integers
\begin{equation}
W_s=\sum_{p=0}^{N_o-1} s_p,
\qquad
W_c=\sum_{p=0}^{N_o-1} c_p
\label{eq:ws_wc}
\end{equation}
onto the registers labeled $W_s$ and $W_c$ in Fig.~\ref{fig:circuit_cmp}. For
a closed-shell reference, $s_p=1$ exactly when the spatial orbital $p$ is
singly occupied in $x$, so at this point the register $W_s$ holds the
seniority $\Omega(x)$. Moreover, $s_p+2c_p=z_{2p}+z_{2p+1}$ for every pair,
so the Hamming weight of $z$ satisfies $W(z)=W_s+2W_c=2e(x)$.

\paragraph{Combiner.}
The combiner $U_W$ assembles the excitation information from the two counts,
acting as
\begin{equation}
U_W\,\ket{W_s}\ket{W_c}\ket{0}_{W}
 = \ket{W_s}\ket{W_c}\ket{W_s+2W_c}_{W},
\label{eq:uw_action}
\end{equation}
and leaves both inputs intact so that the counting networks can later be
uncomputed. Since the output register has $\lceil\log_2(n{+}1)\rceil$ bits,
$U_W$ is built from $\mathcal{O}(\log n)$ of the same full adders as
$U_{\mathrm{cnt}}$ and uses $\mathcal{O}(\log n)$ ANDs.

\paragraph{Phase flip.}
After $U_W$, the register $W$ holds $2e(x)$ and the register $W_s$ holds
$\Omega(x)$, and the phase flip acts diagonally on these two registers as
\begin{equation}
Z_{\mathcal{R}}\,\ket{w}_{W}\ket{u}_{W_s}
 = (-1)^{[(w/2,\,u)\in\mathcal{R}]}\,
   \ket{w}_{W}\ket{u}_{W_s},
\label{eq:zr_action}
\end{equation}
where $[\cdot]$ denotes the indicator function. It can be implemented as a
product of one multi-controlled $Z$ gate per accepted pair $(e,\Omega)$, each
acting on $\mathcal{O}(\log n)$ controls, which adds only subleading T gates.

\paragraph{T-gate count.}
Collecting the contributions,
\begin{align}
N_T^{e,\Omega}
  &= \underbrace{4\cdot\tfrac{n}{2}}_{U_{\mathrm{HA}}}
   + \underbrace{4\bigl(\tfrac{n}{2}-\mathrm{wt}_2(\tfrac{n}{2})\bigr)}_{U_{\mathrm{cnt}}\ \text{on}\ \{s_p\}}
   \nonumber\\
  &\quad
   + \underbrace{4\bigl(\tfrac{n}{2}-\mathrm{wt}_2(\tfrac{n}{2})\bigr)}_{U_{\mathrm{cnt}}\ \text{on}\ \{c_p\}}
   + \underbrace{\mathcal{O}(\log n)}_{U_W,\ Z_{\mathcal{R}}}
   \nonumber\\
  &\sim 6n,
\label{eq:tcount_general_app}
\end{align}
and the uncomputation of all three networks is free. The $2n$ excess over
the threshold variants is the price of holding $e$ and $\Omega$ in two
independent registers rather than computing a single combined quantity.

\paragraph{Open-shell reference.}
For an open-shell reference, the sum bit of a singly occupied reference pair
indicates the absence rather than the presence of seniority in $x$, so $W_s$
no longer equals $\Omega(x)$, and unlike in the hierarchy case the mismatch
cannot be absorbed into the phase flip. In this case $\Omega(x)$ is obtained
as the Hamming weight of the corrected bits $s_p\oplus o_p$, where
$o_p=1$ on the singly occupied reference pairs, written by the same CNOTs
together with classically placed $X$ gates at no T cost, and $W(z)=2e(x)$ is
obtained by a direct Hamming-weight computation on the $n$ bits of $z$. The
two computations
cost $4\bigl(N_o-\mathrm{wt}_2(N_o)\bigr)$ and
$4\bigl(n-\mathrm{wt}_2(n)\bigr)$ T gates, no carry bits are needed, and the
total remains bounded by $6n$, so the leading cost is unchanged.

\FloatBarrier
\bibliographystyle{apsrev4-2}
\bibliography{bib}

@misc{Kitaev1995,
  title = {Quantum measurements and the {Abelian} stabilizer problem},
  author = {Kitaev, A. Yu.},
  year = {1995},
  eprint = {quant-ph/9511026},
  archivePrefix = {arXiv}
}

@article{AspuruGuzik2005,
  title = {Simulated Quantum Computation of Molecular Energies},
  author = {Aspuru-Guzik, Al{\'a}n and Dutoi, Anthony D. and Love, Peter J. and Head-Gordon, Martin},
  journal = {Science},
  volume = {309},
  number = {5741},
  pages = {1704--1707},
  year = {2005},
  doi = {10.1126/science.1113479}
}

@article{Reiher2017,
  title = {Elucidating reaction mechanisms on quantum computers},
  author = {Reiher, Markus and Wiebe, Nathan and Svore, Krysta M. and Wecker, Dave and Troyer, Matthias},
  journal = {Proc. Natl. Acad. Sci. U.S.A.},
  volume = {114},
  number = {29},
  pages = {7555--7560},
  year = {2017},
  doi = {10.1073/pnas.1619152114}
}

@article{Lee2021,
  title = {Even More Efficient Quantum Computations of Chemistry Through Tensor Hypercontraction},
  author = {Lee, Joonho and Berry, Dominic W. and Gidney, Craig and Huggins, William J. and McClean, Jarrod R. and Wiebe, Nathan and Babbush, Ryan},
  journal = {PRX Quantum},
  volume = {2},
  pages = {030305},
  year = {2021},
  doi = {10.1103/PRXQuantum.2.030305}
}

@article{QSCI,
  title = {Quantum-selected configuration interaction: Classical diagonalization of {Hamiltonians} in subspaces selected by quantum computers},
  author = {Kanno, Keita and Kohda, Masaya and Imai, Ryosuke and Koh, Sho and Mitarai, Kosuke and Mizukami, Wataru and Nakagawa, Yuya O.},
  journal = {Phys. Rev. Res.},
  volume = {8},
  pages = {023268},
  year = {2026},
  publisher = {American Physical Society},
  doi = {10.1103/dmn4-snfx}
}

@article{ADAPT-QSCI,
  title = {ADAPT-QSCI: Adaptive Construction of an Input State for Quantum-Selected Configuration Interaction},
  author = {Nakagawa, Yuya O. and Kamoshita, Masahiko and Mizukami, Wataru and Sudo, Shotaro and Ohnishi, Yu-ya},
  journal = {Journal of Chemical Theory and Computation},
  volume = {20},
  number = {24},
  pages = {10817--10825},
  year = {2024},
  publisher = {American Chemical Society},
  doi = {10.1021/acs.jctc.4c00846},
  url = {https://pubs.acs.org/doi/10.1021/acs.jctc.4c00846}
}

@article{TE-QSCI,
  title = {Quantum-selected configuration interaction with time-evolved state},
  author = {Mikkelsen, Mathias and Nakagawa, Yuya O.},
  journal = {Phys. Rev. Research},
  volume = {7},
  pages = {043043},
  year = {2025},
  publisher = {American Physical Society},
  doi = {10.1103/75pv-hbrx}
}

@article{HSB-QSCI,
  title = {Hamiltonian simulation-based quantum-selected configuration interaction for large-scale electronic structure calculations with a quantum computer},
  author = {Sugisaki, Kenji and Kanno, Shu and Itoko, Toshinari and Sakuma, Rei and Yamamoto, Naoki},
  journal = {Phys. Chem. Chem. Phys.},
  volume = {27},
  pages = {20869--20884},
  year = {2025},
  publisher = {Royal Society of Chemistry},
  doi = {10.1039/D5CP02202A}
}

@article{QSCI-IBM,
  title = {Chemistry beyond the scale of exact diagonalization on a quantum-centric supercomputer},
  author = {Robledo-Moreno, Javier and Motta, Mario and Haas, Holger and Javadi-Abhari, Ali and Jurcevic, Petar and Kirby, William and Martiel, Simon and Sharma, Kunal and Sharma, Sandeep and Shirakawa, Tomonori and Sitdikov, Iskandar and Sun, Rong-Yang and Sung, Kevin J. and Takita, Maika and Tran, Minh C. and Yunoki, Seiji and Mezzacapo, Antonio},
  journal = {Science Advances},
  volume = {11},
  number = {25},
  pages = {eadu9991},
  year = {2025},
  publisher = {AAAS},
  doi = {10.1126/sciadv.adu9991}
}

@misc{AFQMC-QSCI,
  title = {Auxiliary-field quantum Monte Carlo method with quantum selected configuration interaction},
  author = {Yoshida, Yuichiro and Erhart, Luca and Murokoshi, Takuma and Nakagawa, Rika and Mori, Chihiro and Miyanaga, Takafumi and Mori, Toshio and Mizukami, Wataru},
  year = {2025},
  eprint = {2502.21081},
  archivePrefix = {arXiv}
}

@article{QSCI-Limitations,
  title = {Critical Limitations in Quantum-Selected Configuration Interaction Methods},
  author = {Reinholdt, Peter and Ziems, Karl Michael and Kjellgren, Erik Rosendahl and Coriani, Sonia and Sauer, Stephan P. A. and Kongsted, Jacob},
  journal = {Journal of Chemical Theory and Computation},
  volume = {21},
  number = {14},
  pages = {6811--6822},
  year = {2025},
  doi = {10.1021/acs.jctc.5c00375}
}

@misc{SQD-AA,
  title = {Sample-Based Quantum Diagonalization with Amplitude Amplification},
  author = {Stockinger, Nina and N{\"u}tzel, Ludwig and Hartmann, Michael J.},
  year = {2026},
  eprint = {2605.02565},
  archivePrefix = {arXiv}
}

@inproceedings{Grover,
  title = {A fast quantum mechanical algorithm for database search},
  author = {Grover, Lov K.},
  booktitle = {Proceedings of the 28th Annual ACM Symposium on Theory of Computing},
  pages = {212--219},
  year = {1996},
  doi = {10.1145/237814.237866}
}

@article{BHMT,
  title = {Quantum amplitude amplification and estimation},
  author = {Brassard, Gilles and H{\o}yer, Peter and Mosca, Michele and Tapp, Alain},
  journal = {Contemporary Mathematics},
  volume = {305},
  pages = {53--74},
  year = {2002},
  doi = {10.1090/conm/305/05215}
}

@article{Bytautas2011,
  title = {Seniority and orbital symmetry as tools for establishing a full configuration interaction hierarchy},
  author = {Bytautas, Laimutis and Henderson, Thomas M. and Jim{\'e}nez-Hoyos, Carlos A. and Ellis, Jason K. and Scuseria, Gustavo E.},
  journal = {The Journal of Chemical Physics},
  volume = {135},
  number = {4},
  pages = {044119},
  year = {2011},
  doi = {10.1063/1.3613706}
}

@article{Bytautas2015,
  title = {Seniority number description of potential energy surfaces: Symmetric dissociation of water, {N$_2$}, {C$_2$}, and {Be$_2$}},
  author = {Bytautas, Laimutis and Scuseria, Gustavo E. and Ruedenberg, Klaus},
  journal = {The Journal of Chemical Physics},
  volume = {143},
  number = {9},
  pages = {094105},
  year = {2015},
  doi = {10.1063/1.4929904}
}

@misc{BBHT,
  title = {Tight bounds on quantum searching},
  author = {Boyer, Michel and Brassard, Gilles and H{\o}yer, Peter and Tapp, Alain},
  year = {1996},
  eprint = {quant-ph/9605034},
  archivePrefix = {arXiv}
}

@article{Kossoski2022,
  title = {Hierarchy Configuration Interaction: Combining Seniority Number and Excitation Degree},
  author = {Kossoski, F{\'a}bris and Damour, Yann and Loos, Pierre-Fran{\c{c}}ois},
  journal = {The Journal of Physical Chemistry Letters},
  volume = {13},
  number = {18},
  pages = {4342--4349},
  year = {2022},
  doi = {10.1021/acs.jpclett.2c00730}
}

@book{Helgaker2000,
  title = {Molecular Electronic-Structure Theory},
  author = {Helgaker, Trygve and J{\o}rgensen, Poul and Olsen, Jeppe},
  publisher = {John Wiley \& Sons},
  address = {Chichester},
  year = {2000},
  isbn = {978-0-471-96755-2},
  doi = {10.1002/9781119019572}
}

@misc{Cuccaro2004,
  title = {A new quantum ripple-carry addition circuit},
  author = {Cuccaro, Steven A. and Draper, Thomas G. and Kutin, Samuel A. and Moulton, David Petrie},
  year = {2004},
  eprint = {quant-ph/0410184},
  archivePrefix = {arXiv}
}

@article{Gidney2018,
  title = {Halving the cost of quantum addition},
  author = {Gidney, Craig},
  journal = {Quantum},
  volume = {2},
  pages = {74},
  year = {2018},
  doi = {10.22331/q-2018-06-18-74}
}

@article{Maslov2016,
  title = {Advantages of using relative-phase {Toffoli} gates with an application to multiple control {Toffoli} optimization},
  author = {Maslov, Dmitri},
  journal = {Physical Review A},
  volume = {93},
  number = {2},
  pages = {022311},
  year = {2016},
  doi = {10.1103/PhysRevA.93.022311}
}

@article{Barenco1995,
  title = {Elementary gates for quantum computation},
  author = {Barenco, Adriano and Bennett, Charles H. and Cleve, Richard and DiVincenzo, David P. and Margolus, Norman and Shor, Peter and Sleator, Tycho and Smolin, John A. and Weinfurter, Harald},
  journal = {Physical Review A},
  volume = {52},
  number = {5},
  pages = {3457--3467},
  year = {1995},
  doi = {10.1103/PhysRevA.52.3457}
}

@article{Matsuzawa2020,
  author = {Matsuzawa, Yuta and Kurashige, Yuki},
  title = {Jastrow-type Decomposition in Quantum Chemistry for Low-Depth Quantum Circuits},
  journal = {Journal of Chemical Theory and Computation},
  volume = {16},
  number = {2},
  pages = {944--952},
  year = {2020},
  doi = {10.1021/acs.jctc.9b00963},
}

@article{Motta2023,
  author = {Motta, Mario and Sung, Kevin J. and Whaley, K. Birgitta and Head-Gordon, Martin and Shee, James},
  title = {Bridging physical intuition and hardware efficiency for correlated electronic states: the local unitary cluster Jastrow ansatz for electronic structure},
  journal = {Chemical Science},
  volume = {14},
  number = {40},
  pages = {11213--11227},
  year = {2023},
  doi = {10.1039/d3sc02516k},
}

@article{Motta2021,
  author = {Motta, Mario and Ye, Erika and McClean, Jarrod R. and Li, Zhendong and Minnich, Austin J. and Babbush, Ryan and Chan, Garnet Kin-Lic},
  title = {Low rank representations for quantum simulation of electronic structure},
  journal = {npj Quantum Information},
  volume = {7},
  pages = {83},
  year = {2021},
  doi = {10.1038/s41534-021-00416-z}
}

@misc{Beverland2022,
  title = {Assessing requirements to scale to practical quantum advantage},
  author = {Beverland, Michael E. and Murali, Prakash and Troyer, Matthias and Svore, Krysta M. and Hoefler, Torsten and Kliuchnikov, Vadym and Low, Guang Hao and Soeken, Mathias and Sundaram, Aarthi and Vaschillo, Alexander},
  year = {2022},
  eprint = {2211.07629},
  archivePrefix = {arXiv}
}

@article{Bocharov2014,
  title = {Efficient Synthesis of Universal Repeat-Until-Success Quantum Circuits},
  author = {Bocharov, Alex and Roetteler, Martin and Svore, Krysta M.},
  journal = {Phys. Rev. Lett.},
  volume = {114},
  issue = {8},
  pages = {080502},
  year = {2015},
  publisher = {American Physical Society},
  doi = {10.1103/PhysRevLett.114.080502}
}

@article{Kivlichan2020,
  title = {Improved Fault-Tolerant Quantum Simulation of Condensed-Phase Correlated Electrons via Trotterization},
  author = {Kivlichan, Ian D. and Gidney, Craig and Berry, Dominic W. and Wiebe, Nathan and McClean, Jarrod and Sun, Wei and Jiang, Zhang and Rubin, Nicholas and Fowler, Austin and Aspuru-Guzik, Al\'an and Neven, Hartmut and Babbush, Ryan},
  journal = {Quantum},
  volume = {4},
  pages = {296},
  year = {2020},
  doi = {10.22331/q-2020-07-16-296}
}

@article{BravyiKitaev2005,
  title = {Universal quantum computation with ideal {Clifford} gates and noisy ancillas},
  author = {Bravyi, Sergey and Kitaev, Alexei},
  journal = {Phys. Rev. A},
  volume = {71},
  number = {2},
  pages = {022316},
  year = {2005},
  publisher = {American Physical Society},
  doi = {10.1103/PhysRevA.71.022316}
}

@article{Fomichev2024,
  title = {Initial state preparation for quantum chemistry on quantum computers},
  author = {Fomichev, Stepan and Hejazi, Kasra and Zini, Modjtaba Shokrian and Kiser, Matthew and Fraxanet, Joana and Casares, Pablo Antonio Moreno and Delgado, Alain and Huh, Joonsuk and Voigt, Arne-Christian and Mueller, Jonathan E. and Arrazola, Juan Miguel},
  journal = {PRX Quantum},
  volume = {5},
  number = {4},
  pages = {040339},
  year = {2024},
  publisher = {American Physical Society},
  doi = {10.1103/PRXQuantum.5.040339}
}

@article{Larsson2022,
  title = {The Chromium Dimer: Closing a Chapter of Quantum Chemistry},
  author = {Larsson, Henrik R. and Zhai, Huanchen and Umrigar, C. J. and Chan, Garnet Kin-Lic},
  journal = {Journal of the American Chemical Society},
  volume = {144},
  number = {35},
  pages = {15932--15937},
  year = {2022},
  doi = {10.1021/jacs.2c06357}
}

@article{Motta2020,
  title = {Ground-State Properties of the Hydrogen Chain: Dimerization, Insulator-to-Metal Transition, and Magnetic Phases},
  author = {Motta, Mario and Genovese, Claudio and Ma, Fengjie and Cui, Zhi-Hao and Sawaya, Randy and Chan, Garnet Kin-Lic and Chepiga, Natalia and Helms, Phillip and Jim{\'e}nez-Hoyos, Carlos A. and Millis, Andrew J. and Ray, Ushnish and Ronca, Enrico and Shi, Hao and Sorella, Sandro and Stoudenmire, Edwin M. and White, Steven R. and Zhang, Shiwei},
  journal = {Phys. Rev. X},
  volume = {10},
  number = {3},
  pages = {031058},
  year = {2020},
  publisher = {American Physical Society},
  doi = {10.1103/PhysRevX.10.031058}
}

@misc{Yu2025SKQD,
  title = {Quantum-Centric Algorithm for Sample-Based {Krylov} Diagonalization},
  author = {Yu, Jeffery and Robledo Moreno, Javier and Iosue, Joseph T. and Bertels, Luke and Claudino, Daniel and Fuller, Bryce and Groszkowski, Peter and Humble, Travis S. and Jurcevic, Petar and Kirby, William and Maier, Thomas A. and Motta, Mario and Pokharel, Bibek and Seif, Alireza and Shehata, Amir and Sung, Kevin J. and Tran, Minh C. and Tripathi, Vinay and Mezzacapo, Antonio and Sharma, Kunal},
  year = {2025},
  eprint = {2501.09702},
  archivePrefix = {arXiv}
}

@misc{Shirakawa2025ClosedLoop,
  title = {Closed-loop calculations of electronic structure on a quantum processor and a classical supercomputer at full scale},
  author = {Shirakawa, Tomonori and Robledo-Moreno, Javier and Itoko, Toshinari and Tripathi, Vinay and Ueda, Kento and Kawashima, Yukio and Broers, Lukas and Kirby, William and Pathak, Himadri and Paik, Hanhee and Tsuji, Miwako and Kodama, Yuetsu and Sato, Mitsuhisa and Evangelinos, Constantinos and Seelam, Seetharami and Walkup, Robert and Yunoki, Seiji and Motta, Mario and Jurcevic, Petar and Horii, Hiroshi and Mezzacapo, Antonio},
  year = {2025},
  eprint = {2511.00224},
  archivePrefix = {arXiv}
}

@article{Yoshioka2025Krylov,
  title = {Krylov diagonalization of large many-body {Hamiltonians} on a quantum processor},
  author = {Yoshioka, Nobuyuki and Amico, Mirko and Kirby, William and Jurcevic, Petar and Dutt, Arkopal and Fuller, Bryce and Garion, Shelly and Haas, Holger and Hamamura, Ikko and Ivrii, Alexander and Majumdar, Ritajit and Minev, Zlatko and Motta, Mario and Pokharel, Bibek and Rivero, Pedro and Sharma, Kunal and Wood, Christopher J. and Javadi-Abhari, Ali and Mezzacapo, Antonio},
  journal = {Nature Communications},
  volume = {16},
  pages = {5014},
  year = {2025},
  doi = {10.1038/s41467-025-59716-z}
}

@misc{GidneyJones2021,
  title = {A {CCCZ} gate performed with 6 {T} gates},
  author = {Gidney, Craig and Jones, N. Cody},
  year = {2021},
  eprint = {2106.11513},
  archivePrefix = {arXiv}
}

@article{JordanWigner1928,
  author  = {Jordan, P. and Wigner, E.},
  title   = {{\"U}ber das Paulische {\"A}quivalenzverbot},
  journal = {Zeitschrift f{\"u}r Physik},
  volume  = {47},
  number  = {9--10},
  pages   = {631--651},
  year    = {1928},
  doi     = {10.1007/BF01331938},
  publisher = {Springer},
}

@misc{Constantinides2025,
  author  = {Constantinides, Nathan and Yu, Jeffery and Devulapalli, Dhruv and
             Fahimniya, Ali and Schaeffer, Luke and Childs, Andrew M. and
             Gullans, Michael J. and Schuckert, Alexander and Gorshkov, Alexey V.},
  title   = {Low-depth fermion routing without ancillas},
  year    = {2025},
  eprint = {2510.05099},
  archivePrefix = {arXiv}
}

@article{BoyarPeralta2008,
  author  = {Boyar, Joan and Peralta, Ren{\'e}},
  title   = {Tight bounds for the multiplicative complexity of symmetric functions},
  journal = {Theoretical Computer Science},
  volume  = {396},
  pages   = {223--246},
  year    = {2008},
  doi     = {10.1016/j.tcs.2008.01.030}
}

@article{KanSymons2025,
  author  = {Kan, Angus and Symons, Benjamin},
  title   = {Resource-optimized fault-tolerant simulation of the {Fermi--Hubbard} model and high-temperature superconductor models},
  journal = {npj Quantum Information},
  volume  = {11},
  pages   = {138},
  year    = {2025},
  doi     = {10.1038/s41534-025-01091-0}
}

@article{MullerPreparata1975,
  author  = {Muller, David E. and Preparata, Franco P.},
  title   = {Bounds to complexities of networks for sorting and for switching},
  journal = {Journal of the ACM},
  volume  = {22},
  pages   = {195--201},
  year    = {1975},
  doi     = {10.1145/321879.321882}
}

@misc{Gidney2024Cultivation,
  title = {Magic state cultivation: growing {T} states as cheap as {CNOT} gates},
  author = {Gidney, Craig and Shutty, Noah and Jones, Cody},
  year = {2024},
  eprint = {2409.17595},
  archivePrefix = {arXiv}
}

\end{document}